\documentclass[12pt,preprint]{aastex}

\shorttitle{Observation and numerical modeling of chromospheric evaporation}
\shortauthors{Imada et al.}

\begin{document}

\title{Observation and numerical modeling of chromospheric evaporation during the impulsive phase of a solar flare}

\author{S. \textsc{Imada},\altaffilmark{1} 
I. \textsc{Murakami},\altaffilmark{2,3}
T.  \textsc{Watanabe},\altaffilmark{4,3}
}
  
\altaffiltext{1}{ Solar-Terrestrial Environment Laboratory (STEL), Nagoya University, Furo-cho, Chikusa-ku, Nagoya 464-8601, Japan}
\altaffiltext{2}{National Institute for Fusion Science}
\altaffiltext{3}{SOKENDAI (The Graduate University for Advanced Studies)}
\altaffiltext{4}{ National Astronomical Observatory of Japan,  2--21--1 Osawa, Mitaka-shi, Tokyo 181--8588, Japan}

\begin{abstract}
We have studied the chromospheric evaporation flow during the impulsive phase of the flare by using the Hinode/EIS observation and 1D hydrodynamic numerical simulation coupled to the time-dependent ionization.
The observation clearly shows that the strong redshift can be observed at the base of the flaring loop only during the impulsive phase.
We performed two different numerical simulations to reproduce the strong downflows in Fe{\small XII} and Fe{\small XV} during the impulsive phase.
By changing the thermal conduction coefficient, we carried out the numerical calculation of chromospheric evaporation in the thermal conduction dominant regime (conductivity coefficient $\kappa_0 = $ classical value) and the enthalpy flux dominant regime  ($\kappa_0 =$ 0.1$\times$classical value).
The chromospheric evaporation calculation in the enthalpy flux dominant regime could reproduce the strong redshift at the base of the flare during the impulsive phase.
This result might indicate that the thermal conduction can be strongly suppressed in some cases of flare. 
We also find that time-dependent ionization effect is importance to reproduce the strong downflows in Fe~{\small XII} and Fe~{\small XV}.
\end{abstract}


\section{INTRODUCTION}

Magnetic reconnection has been discussed as one of the important mechanisms for the rapid energy conversion of stored free magnetic energy to plasma energy. 
One of the most famous phenomena associated with magnetic reconnection is the solar flare. 
A solar flare is a sudden brightening observed in almost all wavelengths. 
The total amount of energy released by a flare is huge and it often reaches 10$^{32}$ erg within an hour. 
Considerable efforts have been devoted toward understanding the physical mechanism of solar flares, and several mechanisms have been proposed over the several decades.
Nowadays, the standard flare model which is based on magnetic reconnection, namely CSHKP model \citep{car,stu,hir,kop}, is widely believed, and modern telescope have confirmed the predicted characteristics from the model (e.g., cusp-like structure in soft X-ray images \citep{tsu}, reconnection inflows \citep{yok}, reconnection outflows (off limb\citep{mck,inn,ima2013}, on disc \citep{har}), plasmoid ejection \citep{ohy,liu}, and Coronal Mass Ejections \citep{sve, ima, ima2011b}). 

Recently, solar atmosphere has been focused as a space laboratory for magnetic reconnection because of its variety in plasma condition. 
Actually, with the solar atmosphere, we can cover from low $\beta$ ($<$ 1) to high $\beta$ ($>$ 1), weakly ionized to fully ionized, and collisional to collisionless plasma.
Observing magnetic reconnection in various plasma conditions is physically interesting in the field of not only solar physics but also other plasma physics.
Especially, in the solar corona, weak Coulomb collisions can affect the plasma dynamics (e.g., thermal conduction, radiative cooling, ionization and recombination). 
In such a plasma condition, thermal conduction have an important role for energy transport during the explosive events.
The plasma dynamics of magnetic reconnection with heat conduction is different from that without heat conduction. 
For example, the adiabatic slow-mode shocks predicted by Petschek \citep{pet1964} might be isothermal shocks by the heat conduction effects \citep{for1989,yok1997}.
Furthermore, the thermal conduction effect can carry the energy to larger scale.
According to the CSHKP model, the released magnetic energy stored in the solar corona is mainly transported to the chromosphere by the thermal conduction and the non-thermal electrons \citep{fis1986}.
With this process the chromospheric plasma is suddenly heated, and a localized high pressure region is formed.
The pressure-gradient force drives the dense plasma to flow up toward the corona along the magnetic fields.
This upflow is called chromospheric evaporation.

The observational studies on evaporation have been done for several decades.
A key piece of evidence for chromospheric evaporation is the detection of fast upflows in a high temperature plasma.
The first observation of chromospheric evaporation was done by the Bragg crystal spectrometer (BCS) on {\it P78-1} \citep{fel1980}, and much work has been done based on the spacecraft observation \citep{ant1982,wat1990,cul1992,ter,mil2009,bro2013,tia2014} such as {\it Solar Maximum Mission, Hinotori,} and {\it Yohkoh}, {\it Solar Heliospheric Observatory}, {\it Hinode}, and {\it Interface Region Imaging Spectrograph}.
Recently, the temperature dependence of chromospheric evaporation is intensively studied.
Inspection of the temperature dependent observation reveals several interesting characteristics of chromospheric evaporation.
For example, the modern observation showed that the temperature of the point separating upflows and downflows (the flow reversal temperature) is located at $\sim$ 1 MK. 
However, some observations show that the flow reversal temperature locates even higher temperature during the impulsive phase \citep{li2011}. 
Recently, to understand these observational facts the numerical modelings are also intensively performed.
Liu {\it et al.} \citep{liu2009} discussed self consistently combined Fokker-Planck modeling of energetic electrons and hydrodynamic simulation of flare.
They compared the results between conductive heating dominant case and direct energetic electron heating dominant case and found that the flow reversal temperature tend to be lower when conductive heating dominates over direct electron heating.
Because the energy input location is different between thermal conduction and energetic electron precipitation, the location of the flow reversal temperature also different in height. 
They also mentioned that the unusual high temperature (2MK) downflow ($\sim 15$ km sec$^{-1}$) without energetic electron is due to the thermal expansion early in the corona.
Brannon \& Longcope  \citep{bra2014} successfully reproduce the observed flow reversal temperature by 1D hydrodynamic simulation omitting gravity effect.
Because they omit gravity, the solar atmosphere is not stratified.
It is not clear whether the observed flow reversal temperature can be reproduced when they take into account the gravity effect. 
The flow reversal temperature becomes one of the hot topics in the field of the solar physics. 
 
\section{OBSERVATION }

On 6 November 2010, Hinode \citep{kos} observed a solar flare (GOES M5.4, peak time 15:27) at the southeast (20$^\circ$ S, 58$^\circ$ E). 
The EUV Imaging Spectrometer (EIS) on board Hinode is a high spectral/spatial resolution spectrometer aimed at studying dynamic phenomena in the corona \citep{cul}. 
The flare we observed were obtained with flare study (HH\_Flare\_180x160\_v2).
EIS performed a coarse raster scan (2 arcseconds slit and 5 arcseconds step with 8 sec exposure: sparse raster) with about 6 minutes cadence over this flaring active region (NOAA 11121) starting before the flare, and the flare occurred during the scan.
EIS data from the raster are processed using the EIS team provided software, which corrects for the flat field, dark current, cosmic rays, hot pixels, and slit tilt.
For thermal reasons, there is an orbital variation of the line position causing an artificial Doppler shift of $\pm$20 km s$^{-1}$ which follows a sinusoidal behavior.
This orbital variation of the line position was corrected using the house keeping data \citep{kam}.
EIS successfully observed the foot point of the flare, where the chromospheric evaporation occurs, during the impulsive phase of the flare.
EIS obtained EUV images and line-of-sight (LOS) velocities estimated by Doppler shift in several emission lines (e.g., Fe~{\small XII}, Fe~{\small XV}, Fe~{\small XXIII}). 
By using these emission lines, we can discuss the temperature dependence of the chromospheric evaporation upflow/condensation downflow during the flare from a few $10^5$ to $10^7$ K.
Simultaneously, the Solar Dynamics Observatory (SDO)/Atmospheric Imaging Assembly (AIA) instrument acquired full-Sun images with a spatial resolution of ~1000 km. 
We use the AIA 193 \AA~ passband to study the temporal evolution of the flare \citep{boe}. 
{\it Reuven Ramaty High Energy Solar Spectroscopic Imager} (RHESSI) also observed the early phase of flare.
The energetic electron ($>$ 25 keV) enhancement was not observed, although the moderate-energy electron ($\sim$ 10 keV) enhancement was observed during the early phase of the flare (not shown here).
The non-thermal electron acceleration might be weak in this event.
Unfortunately, RHESSI could not observe the main phase of flare because of night time.

Figure 1 shows the flare image of AIA 193 \AA~ channel at the impulsive (1a) and the peak (1b) phase.
The light curve of GOES 1.0-8.0 and 0.5-4.0 \AA~ are also shown in Figure 1c.
The dashed vertical lines show the acquired time of the AIA images (a: impulsive, b: peak).
During the impulsive phase, we can clearly see the two bright points at $(X, Y) \sim (-775'', -370'')$ and $(-775'', -355'')$. 
It seems that these bright points represents the two footpoints of the flaring loop.
Figure 1a represents the very beginning of the flare, because the bright points are limited in space.
We can clearly observe the flare arcades at the peak time of the GOES light curve in Figure 1b, $\sim$4 minutes later from Figure 1a.
There are no clear bright points in Figure 1b any more.

Figure 2 shows the intensity (top: a-c) and LOS velocity (bottom: d-f) images of Fe~{\small XII} ($10^{6.2}$K), Fe~{\small XV} ($10^{6.4}$K), and Fe~{\small XXIII} ($10^{7.2}$K) during the impulsive phase of the flare.
The slit scanning starts from 15:28:08 UT, and the slit is located at the flaring region during the impulsive phase.
We can also see the bright points in Figure 2a-b, which we already discussed in Figure 1a. 
On the other hand, Fe~{\small XXIII} intensity image, which generally represents the hot flaring plasma, shows the flare loop which connects two brights in Figure 2a-b.
This result indicates that the bright points certainly locate at the footpoints of the flaring loop.
The LOS velocity in Fe~{\small XII} at the footpoint of the flaring loop during the impulsive phase shows a few 10 km sec$^{-1}$ downflow. 
The Fe~{\small XV} Doppler image (Figure 2d) also shows that the redshift reaches 50 km sec$^{-1}$.
On the other hand, the Fe~{\small XXIII} Doppler image (Figure 2f) shows the $\sim$ 50 km sec$^{-1}$ upflow.
We can see the temperature dependent flows of the chromospheric evaporation.

From Figure 2, we find the dependence of flow on temperature.
To understand this dependence in detail, we analyze the line profiles more carefully.
Figure 3a-g show the line profiles of O~{\small VI} (184.12 \AA, $10^{5.5}$K), Fe~{\small X} (184.54 \AA, $10^{6.0}$K), Fe~{\small XII} (192.39 \AA, $10^{6.1}$K), Fe~{\small XIV} (264.79 \AA, $10^{6.3}$K), Fe~{\small XV} (284.16 \AA, $10^{6.3}$K), Fe~{\small XVI}(262.98 \AA, $10^{6.4}$K), and Fe~{\small XXIII} (263.77 \AA, $10^{7.1}$K) with a single Gaussian fitting, respectively.
The line profiles in Figure 3 are obtained at the x-mark position in Figure 2.
The dashed lines show the line center wavelength of stationary component for each emission lines.
The dotted lines represent the line center wavelength estimated by single gaussian fitting.
We can clearly see that the all line profiles are redshifted (dashed line $<$ dotted line) except Fe~{\small XXIII} (Figure 3g).
Figure 4 shows the summary of temperature dependence of the chromospheric evaporation upflow/condensation downflow during the impulsive phase of the flare.
Interestingly, the strongest redshift is observed in Fe~{\small XV} emission line which generally represents the hot loops of the active region. 

Figure 5 shows the EIS observation of the peak phase of the flare.
We cannot see the bright points anymore, and the well-developed flare arcades are clearly observed.
The blueshift component in Fe~{\small XXIII} (arrow in Figure 5c) which observed the boundary of the flare arcade might be related to the evaporation upflow.  
The strong redshift cannot be observed in Fe~{\small XII}  or Fe~{\small XV} any more.
Therefore, we think that the strong redshift in hot emission line can be observed  at the footpoint of the flare loop only during the impulsive phase.

\section{Numerical modeling} 

We have shown the Hinode/EIS observation of chromospheric evaporation upflow and condensation downflow during the flare.
The observation shows that we can observe the strong downflows of hot plasma (a few MK) during the impulsive phase of the flare.
In this section, we carry out two numerical modelings (case1: thermal conduction dominant, case2: enthalpy flux dominant) in order to understand the observed strong downflow during the impulsive phase.
The calculations discussed  in this paper were carried out using the 1D version of the numerical package CANS (Coordinated Astronomical Numerical Software; http://www-space.eps.s.u-tokyo.ac.jp/~yokoyama/etc/cans/).

Recently, Imada {\it et al.} \citep{ima2011a,ima2011b}  pointed out that ionization cannot reach equilibrium in some part of flaring region because of its fast flow and rapid heating. 
It is important to take into account time-dependent ionization process when we interpret the observation of flare associated phenomena.  
Thus, we have combined the time-dependent ionization code  with 1D version of the numerical package.

\subsection{Simulation setup}
The simulation setup is almost the same as Imada \& Zweibel \citep{ima2012}.
A fixed semi circular single magnetic loop with a constant cross section is assumed.
We assume the half-length of the loop (L) is  26 Mm.
The dynamics in only half of the loop using a 1D single fluid hydrodynamic (HD) code are calculated, because the symmetry about the loop top can be assumed.
We used 2001 grid points in x. 
Grid spacing in the corona $(x>1.3 x_{tr})$ is set to be $\Delta x_{i+1}= 1.02 \Delta x_{i}$, where $x_{tr}$ is the transition region height ($x_{tr} = 2500$ km).
Below the transition region $(x < 1.3x_{tr})$ we use $\Delta x=$ 0.01 $h_0$, where $h_0$ is the pressure scale height in the chromosphere ($h_{0} = 200$ km).
The reflecting boundary conditions at x=0 and L; $\partial \rho/\partial x=0$, $\partial p /\partial x=0$, $V_x=0$ are used.

The 1D hydrodynamic equation in Eulerian form are
\begin{equation}
\frac{\partial \rho}{\partial t} + \frac{\partial}{\partial x}\left(\rho V_x \right)=0,
\end{equation}
\begin{equation}
\frac{\partial}{\partial t}\left(\rho V_x \right) + \frac{\partial}{\partial x}\left(\rho V_x^2 +p \right) =-\rho g_{\parallel} ,
\end{equation}
\begin{equation}\label{energy}
\frac{\partial}{\partial t}\left(\frac{p}{\gamma-1} + \frac{1}{2}\rho V_x^2\right)  + \frac{\partial}{\partial x}\left[\left(\frac{\gamma}{\gamma-1}p+\frac{1}{2}\rho V_x^2\right) V_x - \kappa_{\parallel} \frac{\partial T}{\partial x} \right]=-\rho g_{\parallel} V_x + H -R,
\end{equation}
\begin{equation}
p=\frac{k_B}{m}\rho T,
\end{equation}
\begin{equation}\label{g}
g_{\parallel}=g_0\cos\left[\left(\pi/2\right)x/L\right],
\end{equation}
where $x$ is the distance along a loop from the bottom of chromosphere, $\rho$ is the mass density,  $v$ is the velocity,  $p$ is the gas pressure, $T$ is the fluid temperature, $m$ is the mean mass per particle ($=0.5m_p$), $g_{\parallel}$ is the solar gravity along the loop, $g_0$ is the solar surface gravity ($2.74\times 10^4$ cm s$^{-2}$), $k_B$ is Boltzmann's constant, and $\gamma$ is the ratio of specific heats for an ideal gas taken to be 5/3. 
We use  the classical conductivity for a fully ionized hydrogen plasma \citep{spit}:
\begin{equation}
\kappa_{\parallel}=\kappa_0 T^{5/2},
\end{equation}
where $\kappa_{0}$ is $9.0 \times10^{-7}$ erg s$^{-1}$ K$^{-1}$ cm$^{-1}$.
The radiative loss rate ($R$) is expressed by
\begin{equation}
R=\rho^2 \lambda_\rho(\rho) \Lambda(T),
\end{equation}
where $\lambda_\rho(\rho)$, $\Lambda(T)$ represent the optical thickness effect on the radiative cooling efficiency and the radiative energy loss function, respectively.
We assume $\lambda_\rho(\rho) = \rho_{cl}/\rho \tanh (\rho / \rho_{cl})$, and $\rho_{cl}= m \times 10^{12}$ g cm$^{-3}$.
As a consequence, radiative cooling is strongly suppressed below the transition region, where the atmosphere is optically thick. 
The radiative energy loss function, $\Lambda(T)$, will be discussed later.

The flare heating is represented by the energy input rate per unit volume (H in Equation 3.3). 
The flare heating is set to be symmetric and placed at the loop top and is assumed to be spatially Gaussian and a step function like in time:
\begin{equation}
H(x,t)= H_{0} q(t) \frac{1}{2\sqrt{2\pi}}\exp\left[ -\frac{(x-L)^2}{2w^2_f} \right] \left[ 1+\tanh \left(\frac{x-20 h_0}{3h_0} \right) \right],
\end{equation}
\begin{equation}
q(t)=\frac{1}{4}\left[1+\tanh\frac{t}{0.1 \tau_0} \right]\left[1-\tanh\frac{t-\tau_f}{0.1 \tau_0} \right],
\end{equation}
where $H_0$, $\tau_f$, $w_f$ represent flare energy input rate (3 erg cm$^{-3}$ s$^{-1}$), flare duration (240 s), flare region (6000 km), respectively.
$\tau_0$ (20 s) is the sound traveling time.

We calculate the time evolution of iron charge states to study the effect of transient ionization in the chromospheric evaporation. 
There are many kinds of atomic species in solar corona, and they mainly radiate line emission in ultra-violet wavelength range by bound-bound process. 
Iron is the most dominant element for radiation in solar coronal plasma (a few MK) \citep{lan2013}. 
Further, the recent space telescopes mainly observe the emission lines from iron. 
Therefore, we concentrate on the calculation of the time-dependent ionization of the iron in this study.
The detail description of time-dependent ionization calculation which is used in this study was well discussed in Imada {\it et al.} \citep{ima2011a}.

 The time-dependent ionization equations for iron is
\begin{equation}
\frac{\partial n^{Fe}_i}{\partial t}+\nabla \cdot n^{Fe}_i {\bf v}  = 
n_e\left[n^{Fe}_{i+1} \alpha^{Fe}_{i+1}+ n^{Fe}_{i-1} S^{Fe}_{i-1}-n_i^{Fe}\left(\alpha^{Fe}_{i}+S^{Fe}_{i}\right)\right],
\end{equation}
where $n_i^{Fe}$ is the number density of the $i$th charge state of the iron.
The radiative and dielectronic recombination coefficients and the collisional ionization coefficients are represented by $\alpha^{Fe}_i$ and $S^{Fe}_i$, respectively. 
All ions and electrons have the same flow speed and temperature in the same location are assumed. 

We have combined 1D HD code with time-dependent ionization calculation code.
Because we assumed that the all ions and electrons have the same temperature and velocity, we can calculate the time-dependent ionization (Equation 3.10) by using the temperature ($T$) and velocity ($V_x$) obtained from 1D HD calculation (Equation 3.1-7).
We calculate the radiative energy loss function by CHIANTI atomic database \citep{lan2013} but with ionic fractions of iron calculated by Equation 3.10.
Generally, bound-bound emissions are heavily affected by the ionic fraction. 
Thus, the radiative cooling may be sensitive to the time-dependent ionization process.
We assumed all the elements except iron are in ionization equilibrium, because the dominant source of radiative energy loss is iron in the solar corona. 
We used the usual coronal abundance \citep{fel} to estimate the line emissions.
Through the radiative energy loss function, the time-dependent ionization process can give feedback to the hydrodynamics.
 
\subsection{Simulation results}
We carry out two kinds of numerical simulations to clarify why the strong downflows can be observed during the impulsive phase.
By changing the thermal conductivity coefficient, we try to study the chromospheric evaporation in the thermal conduction dominant regime and in the enthalpy flux dominant regime.

\subsubsection{Case1:  Thermal conduction dominant regime}
First we show the result of chromospheric evaporation calculation in the thermal conduction dominant regime.
As mentioned in Section 1, the solar corona is believed to be in this regime. 
In this calculation, we have used the classical thermal conduction coefficient. 

We assume the loop is initially in hydrostatic equilibrium.
The temperature along the coronal loop is set to be as follows;
\begin{equation}
T(x)=T_0+\frac{1}{2}\left(T_{top}-T_0 \right)\left(\tanh \left(\frac{x-x_{tr}}{0.5h_0}\right)+1 \right),
\end{equation}  
where $T_0$, $T_{top}$ are the temperature of chromospheric plasma  ($10^4$ K) and loop top (2 MK), respectively. 
Our initial condition does not satisfy thermal equilibrium.
However, this is not sensitive to the result of our calculation, because the plasma dynamically changes by the energy input from the flare.
We also assume the ionization equilibrium for initial condition everywhere. 
Figure 6 shows the initial condition of the calculation.
The horizontal axis (x) shows the height along the loop from the bottom of chromosphere. 
The chromosphere locates $0 <x <2$, and the corona locates $2.5 <x$, respectively.
From the top, plasma density (6a), velocity (6b), and temperature (6c) are shown, respectively. 
The intensity, $\tilde{I}$, of Fe~{\small XII} (6d), Fe~{\small XV} (6e), Fe~{\small XXIII} (6f) in each height are also shown, respectively.
The intensities are calculated by CHIANTI atomic database by using the electron density, temperature and $n_i^{Fe}$ and  are normalized by the typical value ($n_i^{Fe}/\Sigma  n_k^{Fe}= 1$ and $n_e = 10^{10}$ cm$^{-3}$).

Figure 7 shows the result of 26 second later from the beginning of the simulation.
From the top, plasma density (7a), velocity (7b), and temperature (7c) are shown, respectively. 
The normalized intensity $\tilde{I}_{Fe~{\small XII}}$ (7d), $\tilde{I}_{Fe~{\small XV}}$  (7f), $\tilde{I}_{Fe~{\small XXIII}}$ (7h) are also shown, respectively.
The black solid line shows the result with time-dependent ionization calculation, and the red dashed lines show the result with ionization equilibrium assumption.
We also show $\tilde{I} \times V_x$ of Fe~{\small XII} (7e), Fe~{\small XV} (7g), Fe~{\small XXIII} (7i) in each height, which represents the Doppler shift contribution parameter in each emission lines. 
The positive value in $\tilde{I} \times V_x$ indicates the blueshift component, and the negative value represents the redshift component. 
$\tilde{I} \times V_x$ are also normalized by the typical value ($V_x=100$km sec$^{-1}$, $n_i^{Fe}/\Sigma  n_k^{Fe}= 1$, and $n_e = 10^{10}$ cm$^{-3}$).

In Figure 7, the thermal conduction front is just reached the chromosphere, and the chromospheric evaporation (Vx $\sim$ 100 km sec$^{-1}$) is forming ($2< x <3$ Mm).
The pressure-driven downflows from the loop top, which carry a large enthalpy flux, do not reach the chromosphere yet.
Its front locates at $x \sim 6$ Mm.
We can also see the condensation downflow (Vx $\sim$ -10 km sec$^{-1}$) at $x \sim 2$ Mm.
At this moment, Fe~{\small XII} locates only at the base of the loop, and Fe{\small XV} is distributed larger region ($2< x <5$ Mm).
On the other hand Fe~{\small XXIII} is not formed yet.
The ionization cannot reach equilibrium in the flaring region because of its fast flow and rapid heating.
Note that the red dashed lines in Figure 7 show the result of ionization equilibrium assumption, and the black solid lines show the result taken into account the time-dependent ionization effect.
Both of the Doppler shift contribution parameter, $\tilde{I} \times V_x$, of  Fe~{\small XII} and Fe~{\small XV} show the positive value, which means that the blueshift component should be observed in these emission lines.

Figure 8 shows the result of 80 second later from the beginning.
The figure format is the same as Figure 7.
At this moment, the chromospheric evaporation flows are well developed, and the highest velocity reaches 250 km sec$^{-1}$.
Although Fe~{\small XII} locates only at the base of the loop, Fe~{\small XV} spread much higher part of the loop  ($2< x <7$ Mm).
The ionization almost reach equilibrium because of the high density condition of the chromospheric evaporation, and we can observe Fe~{\small XXIII} in most of higher part of the loop.
All of the Doppler shift contribution parameter, $\tilde{I} \times V_x$, of  Fe~{\small XII}, Fe~{\small XV}, and Fe~{\small XXIII} also show the positive (blueshift) value.
Therefore, the observed strong downflow in Fe~{\small XII} and Fe~{\small XV}  emission line during the impulsive phase cannot be reproduced by the chromospheric evaporation calculation in the thermal conduction dominant regime.

\subsubsection{Case2:  Enthalpy flux dominant regime}
In the previous subsection we clearly show that the chromospheric evaporation calculation in the thermal conduction dominant regime cannot reproduce the observed strong downflow in Fe~{\small XII} and Fe~{\small XV}  emission line during the impulsive phase.
Therefore, we change the thermal conduction coefficient to simulate the chromospheric evaporation in the enthalpy flux dominant regime. 
In this calculation we use 10\% of the classical value of the thermal conduction coefficient.

Figure 9 shows the result of 40 second later from the beginning.
The figure format is the same as Figure 7.
The pressure-driven downflows from the loop top, which carry a large enthalpy flux, just reach the chromosphere.
Because of the small thermal conduction coefficient,  in this regime the enthalpy flux carry much more energy than the thermal conduction. 
The highest velocity of pressure-driven downflows almost reaches 150 km sec$^{-1}$.
At this moment, Fe~{\small XII} locates only at the base of the loop, and Fe~{\small XV} is distributed slightly higher region ($2< x <3$ Mm).
We can observe Fe~{\small XXIII} in most of higher part of the loop.
Surprisingly, in Figure 9, both of the Doppler shift contribution parameter, $\tilde{I} \times V_x$, of  Fe~{\small XII} and Fe~{\small XV} show the negative value, which means that the redshift component should be observed in these emission lines.
The redshift component is stronger when the time-dependent ionization are taken into account.

Figure 10 shows the result of 80 second later from the beginning.
The figure format is the same as Figure 7.
Even in this enthalpy flux dominant regime, the chromospheric evaporation flows are well developed, and the highest velocity also reaches 250 km sec$^{-1}$.
Although Fe~{\small XXIII} distribution is slightly different, the result in the enthalpy flux dominant regime is almost the same as the result from that in the case of thermal conduction dominant regime.
The Doppler shift contribution parameters of  Fe~{\small XII} and Fe~{\small XV} show both positive and negative value, although that of Fe~{\small XXIII} show the positive value.
Therefore, the strong downflow in Fe~{\small XII} and Fe~{\small XV} can be observed only the beginning of the flare at the base of flaring loop.

\section{Summary \& Discusiion} 

We have studied the chromospheric evaporation flow during the impulsive phase of the flare by using the Hinode/EIS observation.
We have also preform the numerical simulation of 1D hydrodynamics with time-dependent ionization.
The observation clearly shows that the strong redshift can be observed at the base of the flaring loop only during the impulsive phase.
We perform the two different numerical simulations to reproduce the strong downflows in Fe~{\small XII} and Fe~{\small XV} during the impulsive phase.
By changing the thermal conduction coefficient, we carry out the numerical calculation of chromospheric evaporation in the thermal conduction dominant regime ($\kappa_0 = $ classical value) and the enthalpy flux dominant regime  ($\kappa_0 =$ 0.1$\times$classical value).
The chromospheric evaporation calculation in the enthalpy flux dominant regime can only reproduce the strong redshift at the base of the flare during the impulsive phase.
This result might indicate that the thermal conduction can be suppressed in some cases of the flare. 
We also find that time-dependent ionization effect is importance to reproduce the strong downflows in Fe~{\small XII} and Fe~{\small XV}.

Let us discuss why the chromospheric evaporation calculation in the thermal conduction dominant regime cannot reproduce the strong redshift in Fe~{\small XII} and Fe~{\small XV}.
The solar atmosphere are gravitationally stratified, and the sharp temperature and density gradient region, so called transition region, is formed in between corona and chromosphere.
In the thermal conduction dominant regime, the released energy by flare is conducted along the field line from loop top to the chromosphere.
The heat from solar corona violates the pressure balance of the gravitationally stratified atmosphere.
The thermal conduction can carry only the energy, and the dense plasma is suddenly heated to form a localized high pressure region.
The localized high pressure region should be formed at the transition region which has a large jump in temperature and density.
This is the reason why the flow reversal temperature locates at the transition region temperature in the thermal conduction regime.
On the other hand, in the case of the enthalpy flux dominant regime, the plasma energy and momentum are mainly transported by pressure-driven downflow from flare region to transition region simultaneously.
Therefore, the bottom part of the flare loop is hit and heated by this downflow.
This process can produce the strong downflow in Fe~{\small XII} and Fe~{\small XV} at the base of flare loop during the impulsive phase.

Finally, we discuss the possibility that the thermal conduction can be suppressed in some cases of the flare. 
The classical thermal conduction in a fully ionized hydrogen plasma is based on the assumption that the mean free path is enough shorter than the characteristic temperature scale height in the solar corona.
In the case that the mean free path becomes comparable to or even greater than the temperature scale height, the normal diffusion approximation for the heat flux breaks down\citep{cow1977, dal1993}.
Furthermore, in the case of solar flare, the large structure of magnetic reconnection can also contribute to suppress the thermal conduction effect.
The loss-cone angle of hot electron in the reconnection region is small because of the large magnetic field gradient between upstream and downstream of reconnection region.
The free streaming electron along the magnetic field from reconnection region to outside might be limited in a collisonless plasma.
Suppression mechanism of thermal conduction during a solar flare is important future work.

\acknowledgments Hinode is a Japanese mission developed and launched by ISAS/JAXA, with NAOJ as a domestic partner and NASA and STFC (UK) as international partners. It is operated by these agencies in co-operation with ESA and NSC (Norway). This work was partially supported by the Grant-in-Aid for Young Scientist B (24740130), by the Grant-in-Aid for Scientific Research B (23340045), by the Grant-in-Aid for Scientific Research B (26287143), by the JSPS Core-to-Core Program (22001), by the JSPS Program for Advancing Strategic International Networks to Accelerate the Circulation of Talented Researchers under grant number G2602.

\begin{figure}
\epsscale{0.7}
\plotone{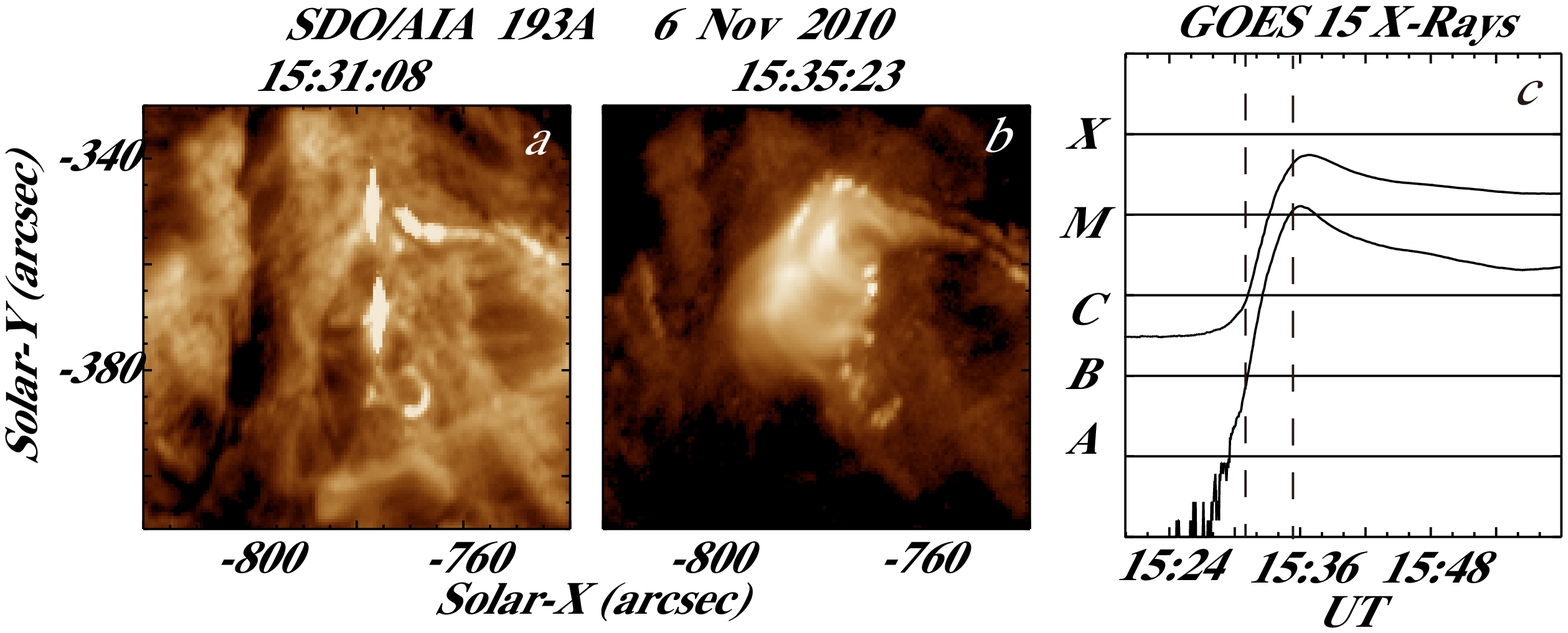}
\caption{The flare image of AIA 193 \AA~ channel at the impulsive (1a) and the peak (1b) phase.
The light curve of GOES 1.0-8.0 and 0.5-4.0 \AA~ are also shown in Figure 1c.
The dashed vertical lines show the acquired time of the AIA images (a: impulsive, b: peak). }
\end{figure}

\begin{figure}
\epsscale{0.9}
\plotone{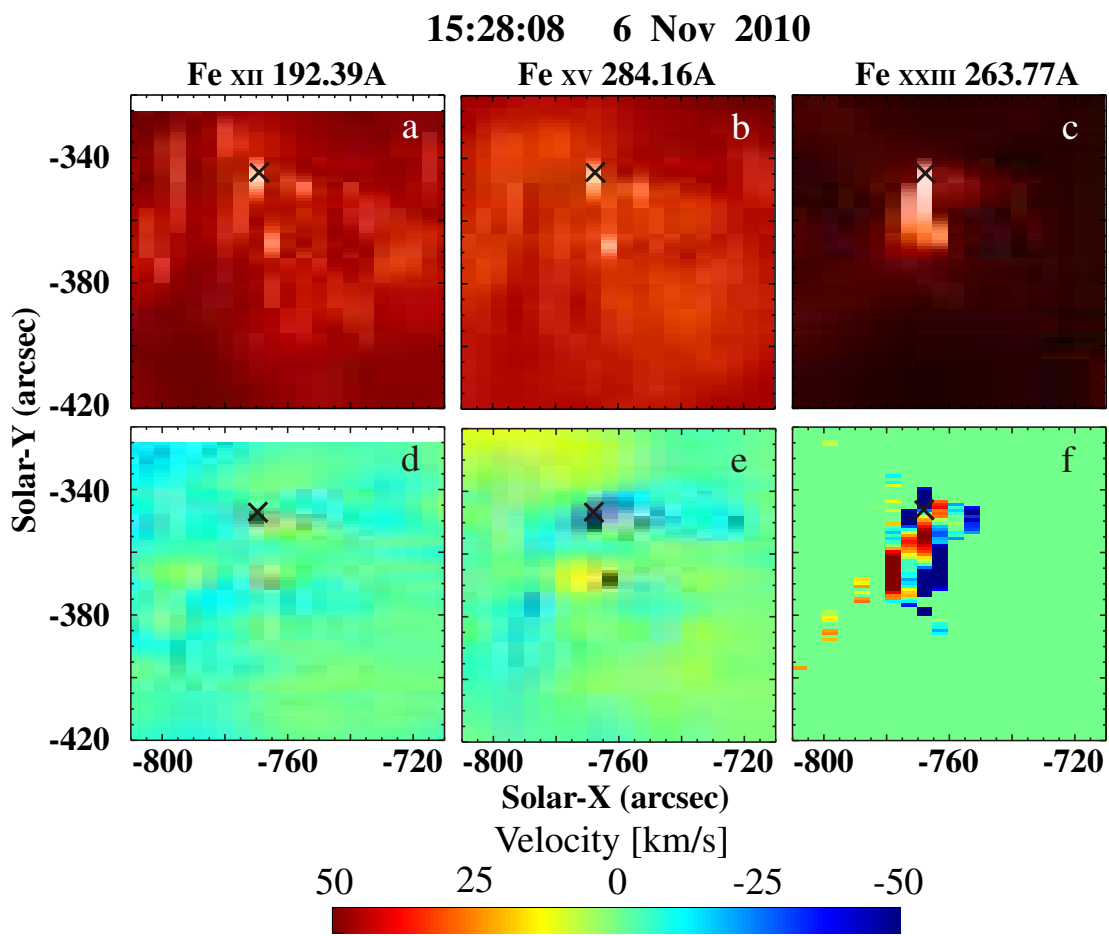}
\caption{The intensity (top: a-c) and LOS velocity (bottom: d-f) images of Fe~{\small XII} ($10^{6.2}$K), Fe~{\small XV} ($10^{6.4}$K), and Fe~{\small XXIII} ($10^{7.2}$K) during the impulsive phase of the flare.}
\end{figure}

\begin{figure}
\epsscale{0.8}
\plotone{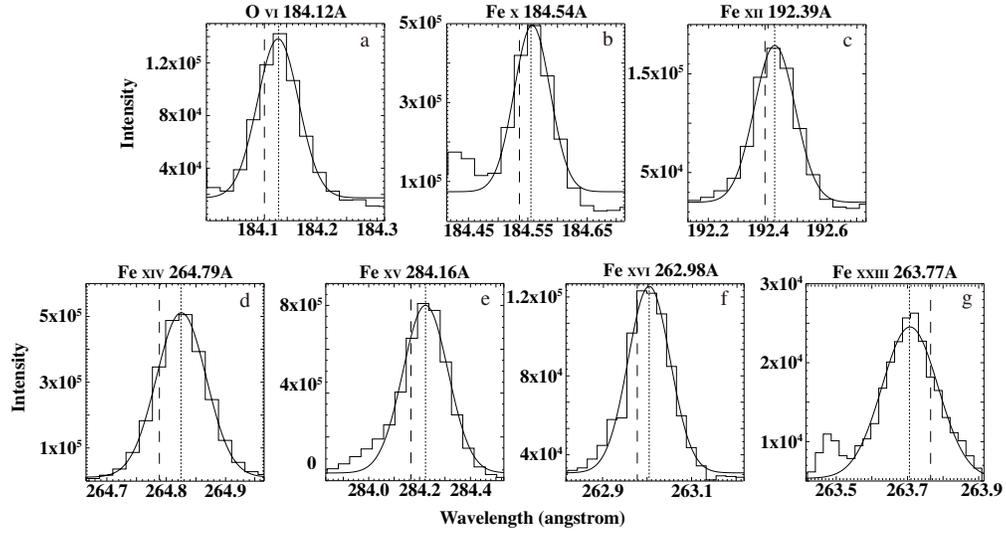}
\caption{The line profile of O~{\small VI} (184.12 \AA, $10^{5.5}$K), Fe~{\small X} (184.54 \AA, $10^{6.0}$K), Fe~{\small XII} (192.39 \AA, $10^{6.1}$K), Fe~{\small XIV} (264.79 \AA, $10^{6.3}$K), Fe~{\small XV} (284.16 \AA, $10^{6.3}$K), Fe~{\small XVI}(262.98 \AA, $10^{6.4}$K), and Fe~{\small XXIII} (263.77 \AA, $10^{7.1}$K) with a single Gaussian fitting.}
\end{figure}

\begin{figure}
\epsscale{0.8}
\plotone{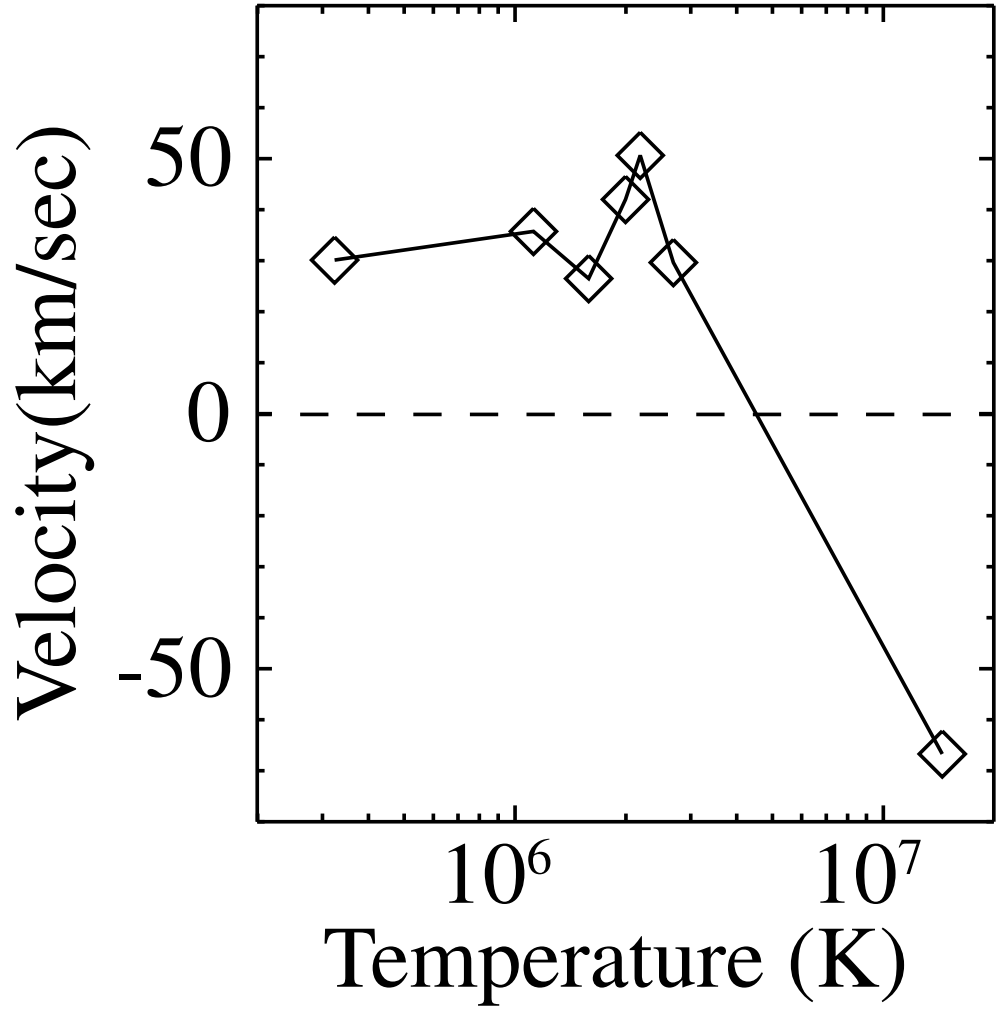}
\caption{The temperature dependence of the chromospheric evaporation upflow/condensation downflow during the impulsive phase of the flare. The LOS velocities are estimated by Doppler shifts.}
\end{figure}

\begin{figure}
\epsscale{0.8}
\plotone{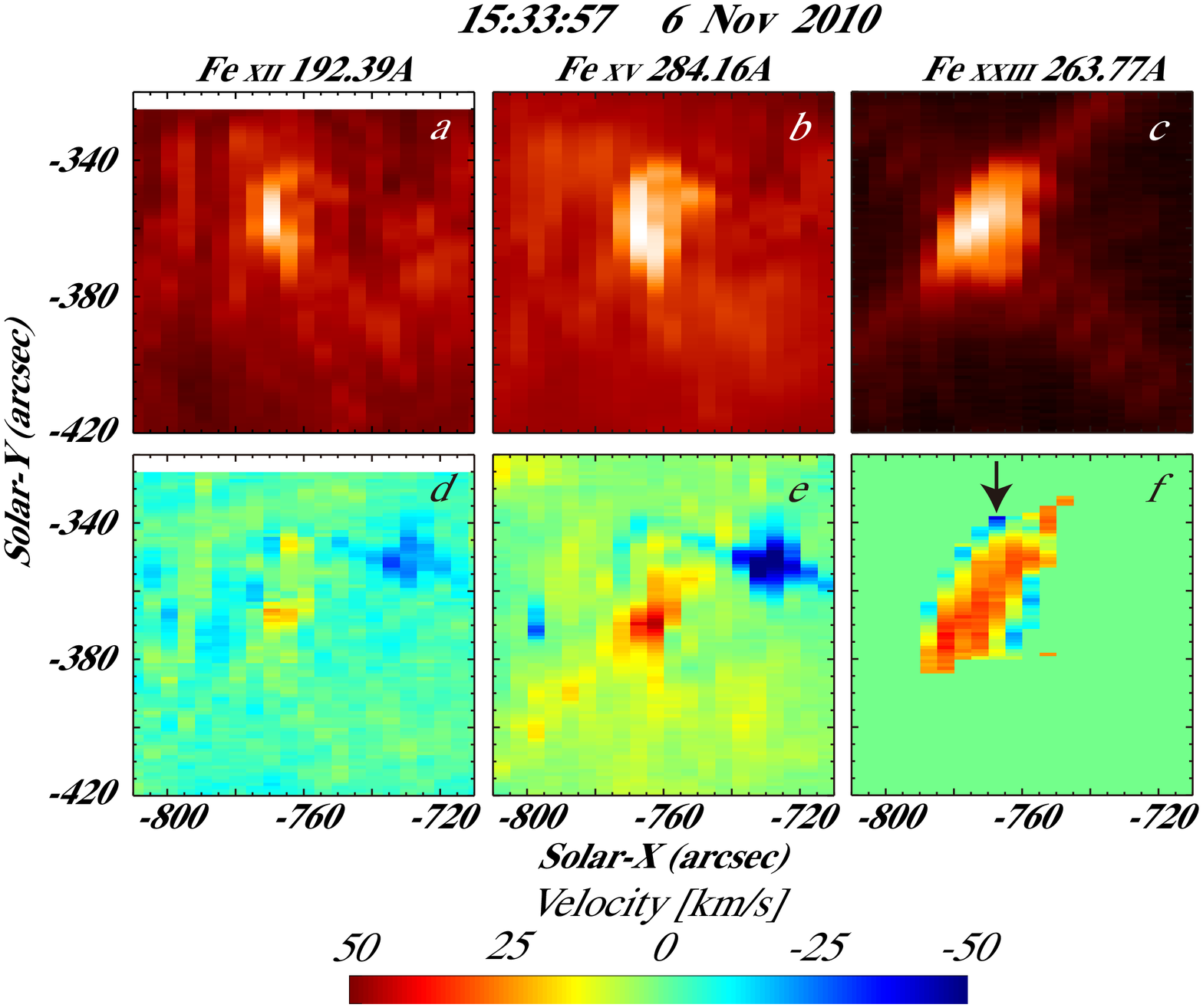}
\caption{The intensity (top: a-c) and LOS velocity (bottom: d-f) images of Fe~{\small XII} ($10^{6.2}$K), Fe~{\small XV} ($10^{6.4}$K), and Fe~{\small XXIII} ($10^{7.2}$K) during the peak phase of the flare.}
\end{figure}

\begin{figure}
\epsscale{0.8}
\plotone{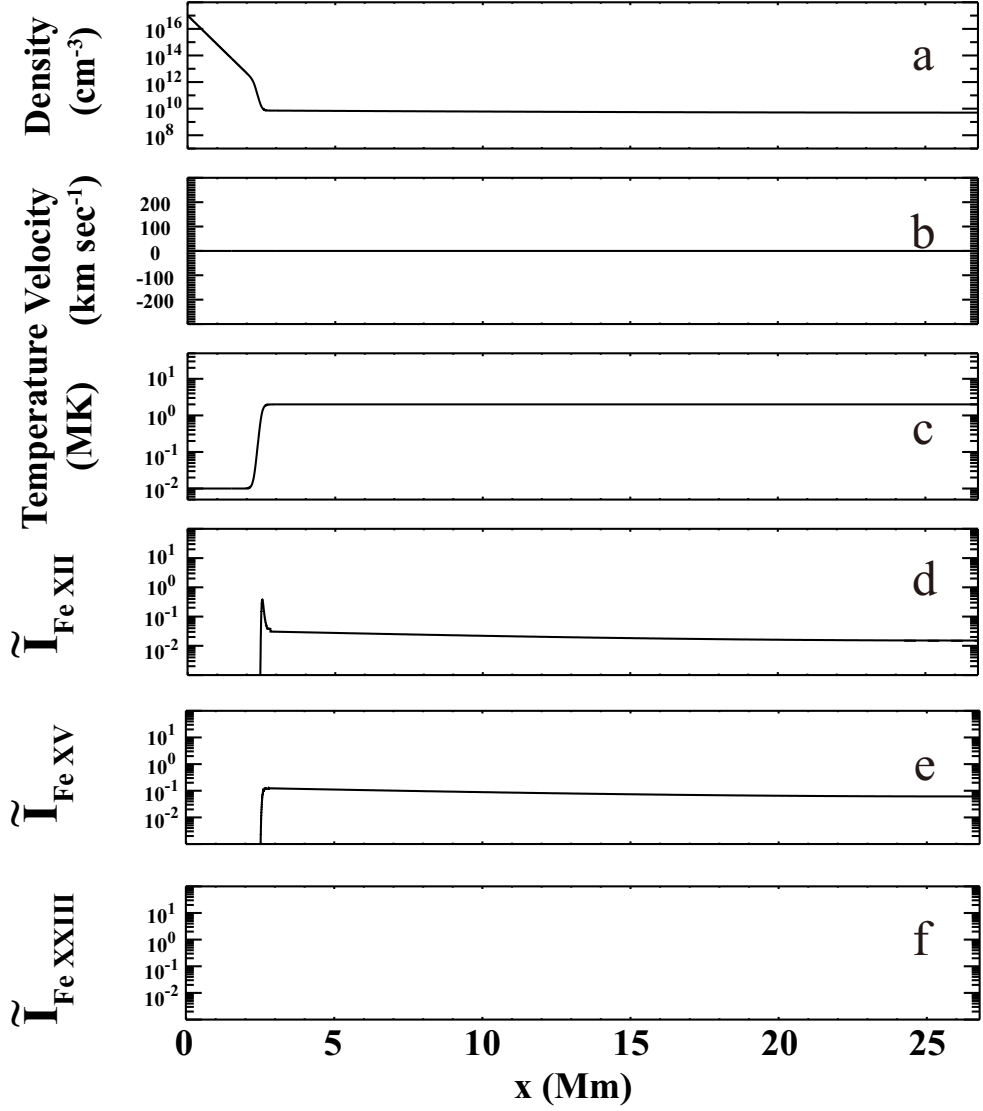}
\caption{The initial condition of the chromospheric evaporation calculation.
From the top, plasma density (6a), velocity (6b), and temperature (6c) are shown, respectively. 
The intensity, $\tilde{I}$, of Fe~{\small XII} (6d), Fe~{\small XV} (6e), Fe~{\small XXIII} (6f) in each height are also shown, respectively.}
\end{figure}

\begin{figure}
\epsscale{0.6}
\plotone{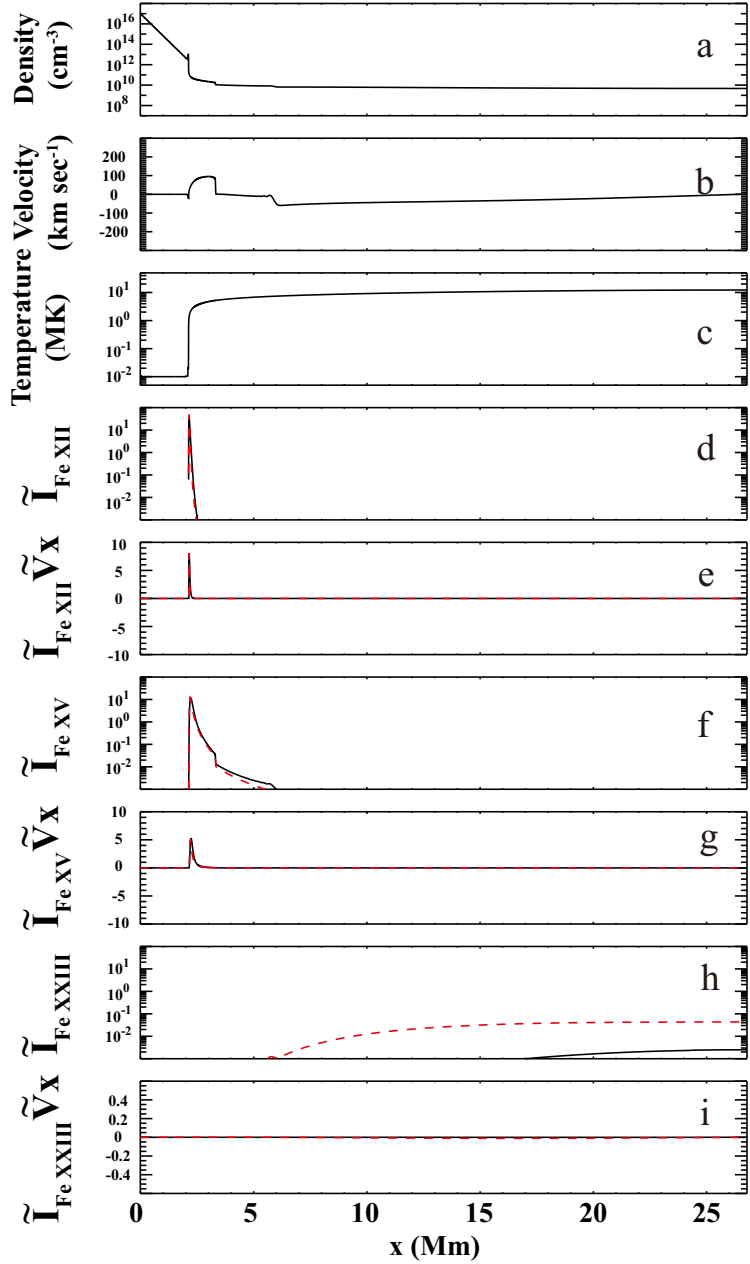}
\caption{The result of 26 second later from the beginning of the simulation in the thermal conduction dominant regime.
The normalized intensity $\tilde{I}_{Fe~{\small XII}}$ (7d), $\tilde{I}_{Fe~{\small XV}}$  (7f), $\tilde{I}_{Fe~{\small XXIII}}$ (7h) are also shown, respectively.
The black solid line shows the result with time-dependent ionization calculation, and the red dashed lines show the result with ionization equilibrium assumption.
We also show $\tilde{I} \times V_x$ of Fe~{\small XII} (7e), Fe~{\small XV} (7g), Fe~{\small XXIII} (7i) in each height, which represents the Doppler shift contribution parameter in each emission lines. }
\end{figure}

\begin{figure}
\epsscale{0.6}
\plotone{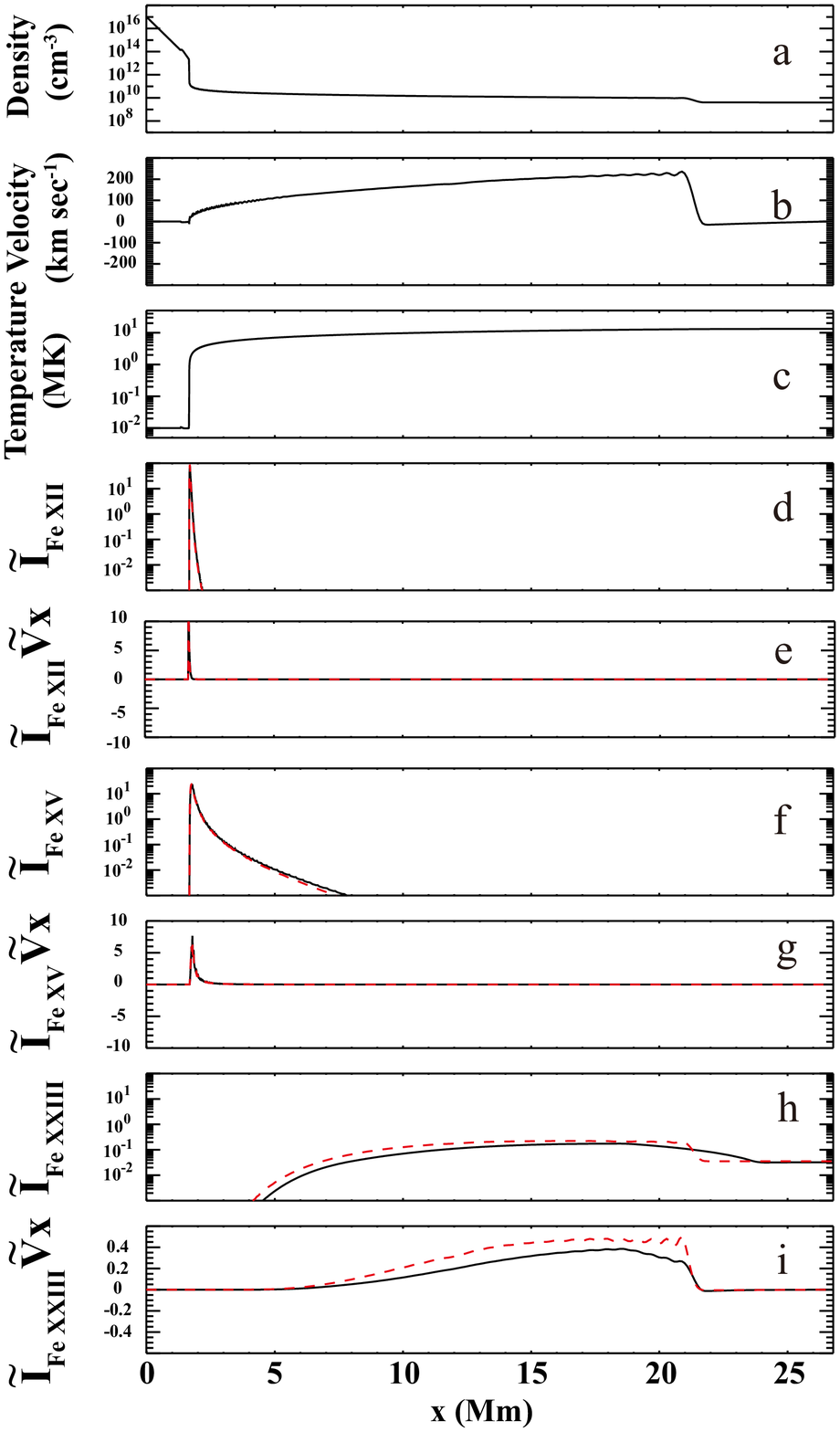}
\caption{The result of 80 second later from the beginning. The figure format is the same as Figure 7.}
\end{figure}

\begin{figure}
\epsscale{0.6}
\plotone{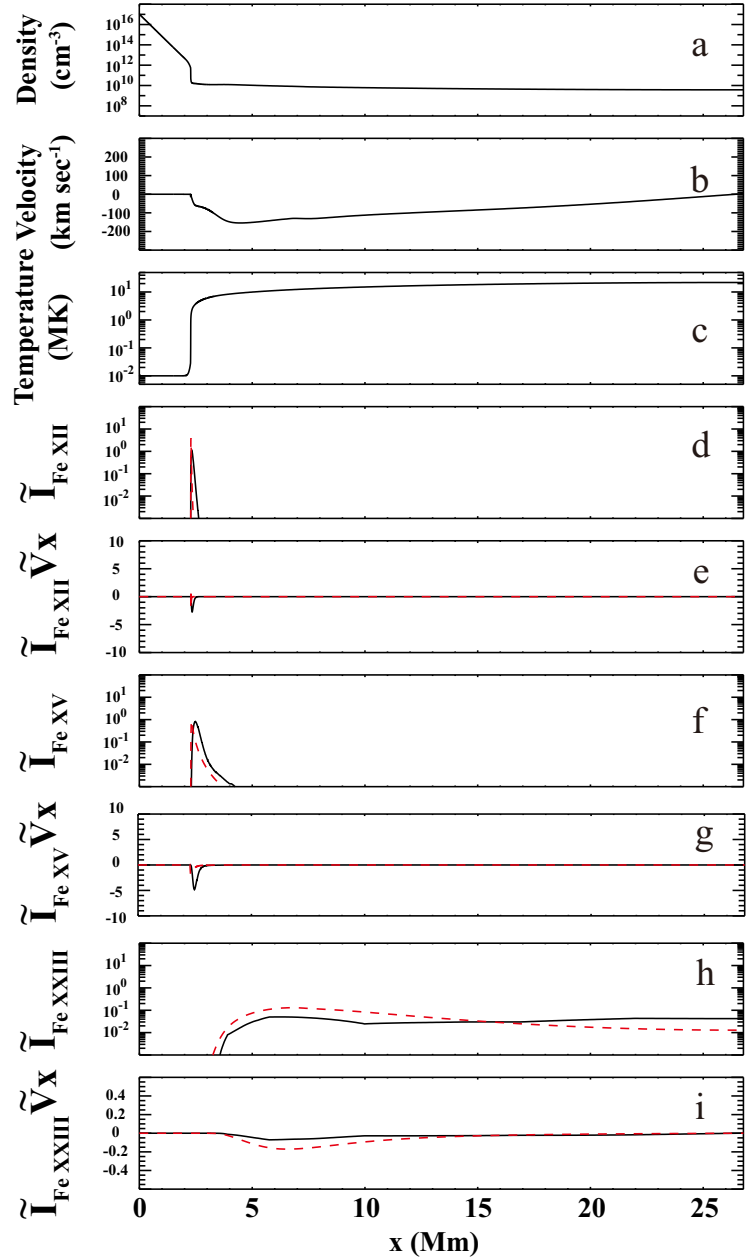}
\caption{The result of 40 second later from the beginning in the enthalpy flux dominant regime.
The figure format is the same as Figure 7.}
\end{figure}

\begin{figure}
\epsscale{0.6}
\plotone{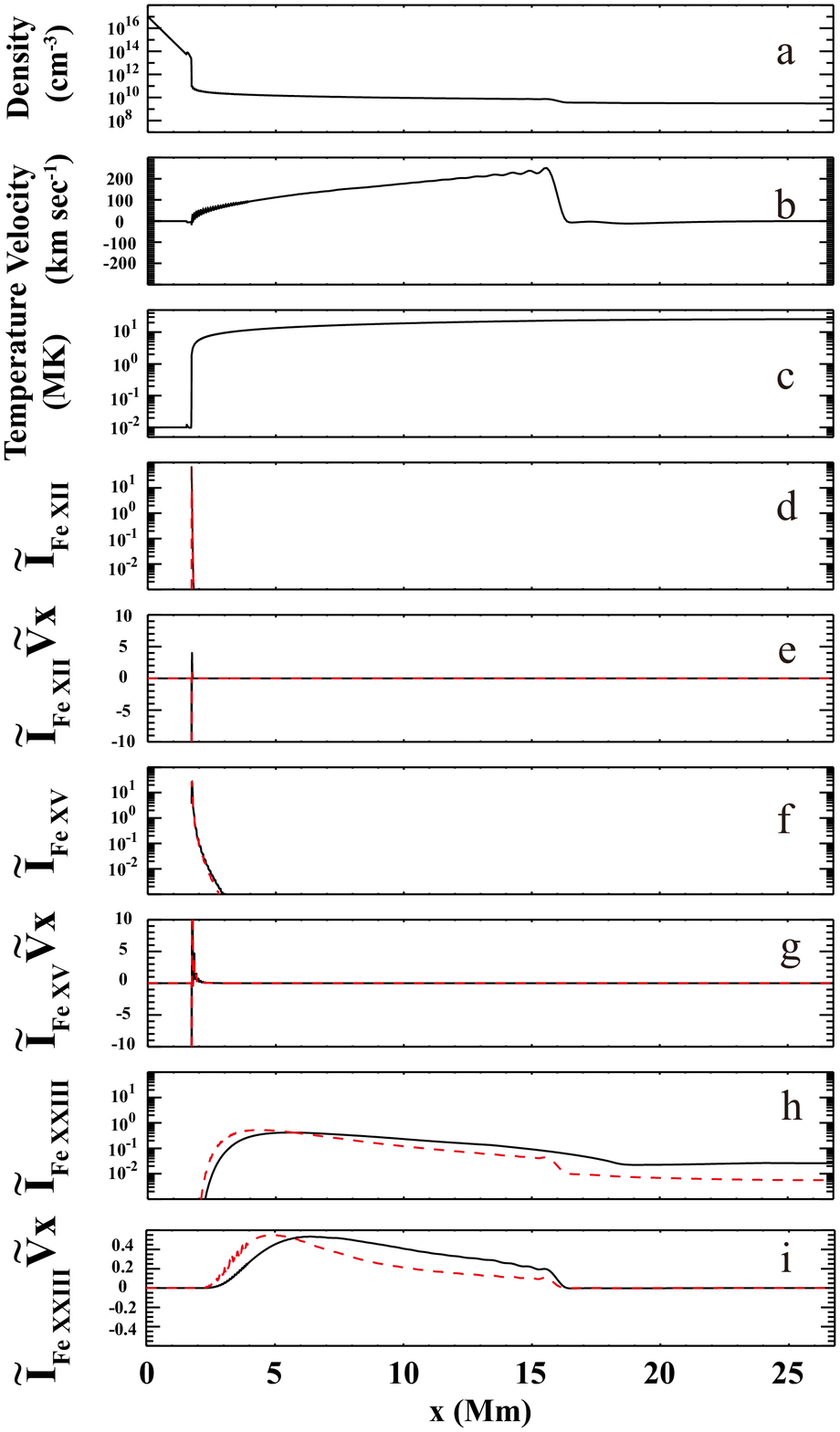}
\caption{The result of 80 second later from the beginning.
The figure format is the same as Figure 7.}
\end{figure}


\begin{thebibliography}{}
\bibitem[Carmichael (1964)]{car} Carmichael, H., {\it in The Physics of Solar Flares, ed. W. N. Hess (NASA Special Publication 50; Washington, DC: NASA)}, 451 (1964).
\bibitem[Sturrock (1966)]{stu}  Sturrock,P. A., Nature, 211, 695 (1966).
\bibitem[Hirayama (1974)]{hir} Hirayama, T., Sol. Phys, 34, 323 (1974). 
\bibitem[Kopp \& Pneuman (1976)]{kop} Kopp, R. A., \& Pneuman, G. W. 1976, Sol. Phys, 50, 85 
\bibitem[Tsuneta et~al. (1992)]{tsu} Tsuneta, S., Hara, H., Shimizu, T., Acton, L. W., Strong, K. T., Hudson, H. S., \& Ogawara, Y. , PASJ, 44, 63 (1992).
\bibitem[Yokoyama et~al. (2001)]{yok} Yokoyama, T., Akita, K., Morimoto, T., Inoue, K., \& Newmark, J., Astrophys. J., 546, L69 (2001).
\bibitem[McKenzie \& Hudson (1999)]{mck} McKenzie, D.~E. \& Hudson, H.~S., Astrophys. J., 519, L93 (1999). 
\bibitem[Innes et~al. (2003)]{inn} Innes, D.~E., McKenzie, D. E., \& Wang, T., Sol. Phys, 217, 267 (2003). 
\bibitem[Imada et~al. (2013)]{ima2013} Imada, S., Aoki, K., Hara, H., Watanabe, T., Harra, L. K., \& Shimizu, T., Astrophys. J., 776, L11 (2013). 
\bibitem[Hara et~al. (2011)]{har} Hara, H., Watanabe, T.,  Harra, L. K., Culhane, J.~L., Young, P. R., Astrophys. J., 741, 107 (2011).
\bibitem[Ohyama \& Shibata (1998)]{ohy} Ohyama, M. \& Shibata, K. 1998, Astrophys. J., 499, 934 (1998). 
\bibitem[Liu et~al. (2013)]{liu} Liu, W. et~al., Astrophys. J., 767, 168 (2013).
\bibitem[Svestka \& Cliver (1992)]{sve} Svestka, Z. \& Cliver, E.~W., History and Basic Characteristics of Eruptive Flares. in IAU Colloq. 133, Eruptive Solar Flares, Vol. 399, ed. Z. Svestka, B. V. Jackson, \& M. E. Macado (New York: Springer), 1 (1992).
\bibitem[Imada et~al. (2007)]{ima} Imada, S., Hara, H., Watanabe, T., Kamio, S., Asai, A., Matsuzaki, K., Harra, L. K., \& Mariska, J. T., PASJ, 59, S793 (2007). 
\bibitem[Imada et~al. (2011b)]{ima2011b} Imada, S., Hara, H., Watanabe, T., Murakami, I., Harra, L. K., Shimizu, T., \& Zweibel, E. G., Astrophys. J., 743, 57 (2011).
\bibitem[Petschek (1964)]{pet1964} Petschek ,~H.E., in NASA Symposium on Physics of Solar Flares (NASA SP 50), 425 (1964).
\bibitem[Forbes et~al. (1989)]{for1989} Forbes, T. G., Malherbe, J. M., \& Priest, E. R., Sol. Phys., 120, 285 (1989).
\bibitem[Yokoyama \& Shibata (1997)]{yok1997} Yokoyama, T. \& Shibata, K., Astrophys. J., 474, L61 (1997).
\bibitem[Fisher (1986)]{fis1986} Fisher, G. H., in IAU Colloq. 89, Radiation Hydrodynamics in Stars and Compact Objects, ed. D. Milhalas \& K.-H. Winkler (Berlin: Springer), 53 (1986).
\bibitem[Feldman et~al. (1980)]{fel1980} Feldman, U., Doschek, G. A., Kreplin, R. W., Mariska, J. T. , Astrophys. J., 241, 1175 (1980).
\bibitem[Antonucci et~al. (1982)]{ant1982} Antonucci, E., Gabriel, A. H., Acton, L. W., Leibacher, J. W., Culhane, J. L., Rapley, C. G., Doyle, J. G., Machado, M. E., \& Orwig, L. E., Sol. Phys, 78, 107 (1982).
\bibitem[Watanabe (1990)]{wat1990} Watanabe, T., Sol. Phys, 126, 351 (1990).
\bibitem[Culhane et~al. (1992)]{cul1992} Culhane, J.~L., Fludra, A., Bentley, R. D., Doschek, G. A., Watanabe, T., Hiei, E., Lang, J., Carter, M. K., Mariska, J. T., Phillips, A. T., Phillips, K. J. H., Pike, C. D., \& Alphones, C., PASJ, 44, L101(1992). 
\bibitem[Teriaca et~al. (2003)]{ter} Teriaca, L., Falchi, A., Cauzzi, G., Falciani, R., Smaldone, L. A., \& Andretta, V., Astrophys. J., 588, 596 (2003).
\bibitem[Milligan \& Dennis (2009)]{mil2009} Milligan, R.~O., \& Dennis, B. R., Astrophys. J., 699, 968 (2009).
\bibitem[Brosius (2013)]{bro2013} Brosius J. W., Astrophys. J., 726, 133 (2013).
\bibitem[Tian et~al. (2014)]{tia2014} Tian, H., Li, G., Reeves, K. K., Raymond, J. C., Guo, F., Liu, W., Chen, B., \& Murphy, N. A., Astrophys. J., 797, L14 (2014).
\bibitem[Li \& Ding (2011)]{li2011} Li, Y., \& Ding, M. D., Astrophys. J., 727, 98 (2011).
\bibitem[Liu et ~al. (2009)]{liu2009} Liu, W., Petrosian, V.,\& Mariska, J. T., Astrophys. J., 702, 1553 (2009).
\bibitem[Brannon \& Longcope (2014)]{bra2014} Brannon S. \& Longcope, D., Astrophys. J., 792, 50 (2014).
\bibitem[Kosugi et~al. (2007)]{kos} Kosugi, T., Matsuzaki, K., Sakao, T., Shimizu, T., Sone, Y., Tachikawa, S., Hashimoto, T., Minesugi, K., Ohnishi, A., Yamada, T., Tsuneta, S., Hara, H., Ichimoto, K., Suematsu, Y., Shimojo, M., Watanabe, T., Shimada, S., Davis, J. M., Hill, L. D., Owens, J. K., Title, A. M., Culhane, J. L., Harra, L. K., Doschek, G. A., \& Golub, L., Sol. Phys, 243, 3 (2007). 
\bibitem[Culhane et~al. (2007)]{cul} Culhane, J.~L., Harra, L. K., James, A. M., Al-Janabi, K., Bradley, L. J., Chaudry, R. A., Rees, K., Tandy, J. A., Thomas, P., Whillock, M. C. R.,  Winter, B., Doschek, G. A., Korendyke, C. M., Brown, C. M., Myers, S., Mariska, J., Seely, J., Lang, J., Kent, B. J., Shaughnessy, B. M., Young, P. R., Simnett, G. M., Castelli, C. M., Mahmoud, S., Mapson-Menard, H., Probyn, B. J., Thomas, R. J., Davila, J., Dere, K., Windt, D., Shea, J., Hagood, R., Moye, R., Hara, H., Watanabe, T., Matsuzaki, K., Kosugi, T., Hansteen, V., \& Wikstol, O., Sol. Phys, 243, 19 (2007).
\bibitem[Kamio et~al. (2010)]{kam} Kamio, S., Hara, H., Watanabe, T., Fredvik, T., Hansteen, V. H., Sol. Phys, 266, 209 (2010).
\bibitem[Boerner et~al. (2012)]{boe} Boerner, P., Edwards, C., Lemen, J., Rausch, A., Schrijver, C., Shine, R., Shing, L., Stern, R., Tarbell, T., Title, A., Wolfson, C. J., Soufli, R., Spiller, E. Gullikson, E., McKenzie, D., Windt, D., Golub, L., Podgorski, W., Testa, P., \& Weber, M., Sol. Phys, 275, 41 (2012).
\bibitem[Imada et~al. (2011a)]{ima2011a} Imada, S., Murakami, I., Watanabe, T., Hara, H., \& Shimizu, T., , Astrophys. J., 742, 70 (2011). 
\bibitem[Imada \& Zweibel (2012)]{ima2012} Imada, S., \& Zweibel, E. G., Astrophys. J., 755, 93 (2012). 
\bibitem[Spitzer (1962)]{spit} Spitzer, L., Physics of Fully Ionizaed Gasses, New York. (1962).
\bibitem[Landi et~al. (2013)]{lan2013} Landi, E., Young, P. R., Dere, K. P., Del Zanna, G., \& Mason, H. E., Astrophys. J., 763, 86 (2013).
\bibitem[Feldman (1992)]{fel} Feldman, U., Phys. Scr., 46,202 (1992).
\bibitem[Cowie \& McKee (1977)]{cow1977} Cowie L. L. \& McKee, C. F., Astrophys. J., 211, 135 (1977).
\bibitem[Dalton \& Balbus (1993)]{dal1993} Dalton W. W. \& Balbus, S. A., Astrophys. J., 404, 625 (1993).
\end{thebibliography}
\end{document}